\documentclass[aps,twocolumn,groupedaddress,showkeys,showpacs]{revtex4}
\usepackage{graphicx}

\begin{document}


\title{A twisted polydiacetylene quantum wire: Influence of conformation on excitons in polymeric quasi-1D systems}
\author{A. Al Choueiry}
\author{T. Barisien}
\email{barisien@insp.jussieu.fr}
\author{J. Holcman}
\author{L. Legrand}
\author{M. Schott}
\affiliation{Institut des NanoSciences de Paris (INSP), UMR 7588/CNRS, Universit\'e Pierre et Marie Curie (Paris 6), Campus Boucicaut, 140 rue de Lourmel, 75015 Paris, France}
\author{G. Weiser}
\affiliation{Faculty of Physics and Centre of Material Science, Philipps-Universit\"{a}t, 35037 Marburg, Germany}
\author{M. Balog}
\author{J. Deschamps}
\author{S.G. Dutremez}
\author{J.-S. Filhol}
\affiliation{Institut Charles Gerhardt Montpellier, UMR 5253/CNRSUM2-ENSCM-UM1, Equipe CMOS, Universit\'e Montpellier II, Place Eug\`ene Bataillon, 34095 Montpellier Cedex 5, France}

\date{\today}

\begin{abstract}
Highly luminescent isolated polydiacetylene (PDA) chains in a new diacetylene crystal are studied. The diacetylene crystal structure and theoretical calculations suggest that the chains are not planar and more strongly twisted than any known PDA. Yet the chains behave as quasi perfect quantum wires, showing that this behaviour is generic among PDA, provided crystal order is high enough. The exciton energy, E$_{0}$, is higher than in any other well ordered PDA providing a relationship between E$_{0}$  and twist angle. On the other hand the exciton binding energy and Bohr radius are significantly affected as demonstrated in electroabsorption experiments (and developed in a companion paper). 
\end{abstract}


\maketitle

\section{Introduction}
Polydiacetylenes (PDA) are a class of conjugated polymers with general formula =(RC-C$\equiv$C-CR')$_{n}$= where the substituents R and R', the so-called side groups, can have very different chemical formulae. They are prepared by topochemical polymerization of the corresponding monomer diacetylene (DA) crystals, suitable side groups ensuring that the geometry is favorable to the solid state reaction \cite{Wegner69, Baughman74, Baughman78, Enkelmann84}.\\
These systems can be used as model conjugated polymers as they can be obtained through the topochemical reaction as quasi-perfectly ordered linear conjugated chains (for a review, see \cite{Schott06a}). In fact, such a chain is a semiconducting quantum wire. The excited state which dominates all optical properties in the visible and near UV is a strongly bound exciton which carries almost all the oscillator strength \cite{Weiser92, Horvath96} and exhibits all theoretical properties expected for a perfect periodic quasi-1D system \cite{Lecuiller02, Dubin02}; moreover, the exciton on these chains has been shown to be in a macroscopically spatially coherent state at low temperature \cite{Dubin06a, Legrand08}. In good enough crystals of the DA 3BCMU and 4BCMU, polariton induced transparency has been observed under low intensity incoherent light without the need of a cavity, thanks to the perfect alignment of the very large transition dipole moments of all polymer chains in the crystal \cite{Weiser07a, Weiser07b}.\\
As a result, PDA chains can be treated as periodic 1D systems, and there are a number of theoretical studies of their electronic structure often on series of oligomers (see \cite{Race01, Race03} and references therein, and \cite{Barcza}). An important issue is to assess the role of electronic correlations in quasi 1D systems.\\
PDA may have at least two different electronic structures, corresponding to the same chemical formula, but presumably to different chain geometries; these geometrical differences have not been clearly identified yet \cite{Schott06b}. The two most frequent structures are conventionally named ``blue'' and ``red'', by the visual color by transmission or diffuse reflection of weakly polymerized powders or crystals. These colors correspond to excitonic absorption near 630 nm and 540 nm, respectively. Thus, the energies of the excitons responsible of that absorption differ by several tenths of an eV.\\
But the most spectacular difference is that, while blue PDA chains are almost not fluorescent (yields 10$^{-4}$ or less), red ones show an intense resonance emission with yields 0.1 or higher at low T \cite{Lecuiller99}. The absence of fluorescence in blue chains corresponds to the presence in the optical gap of ``dark'' exciton states of \textit{g} symmetry \cite{Lawrence94, Kobayashi96, Kraabel98a} which provide a very efficient non radiative decay channel for the exciton states responsible for light absorption and emission. By implication, the dark states in the red chains are no longer in the optical gap, so their energies have increased much more than that of the radiating exciton on passing from blue to red chains.\\
Crystal structure determinations have shown beyond doubt that blue chains are planar and that successive repeat units are translationally equivalent: the 1D unit cell contains a single repeat unit \cite{Enkelmann84, Schott87}. It has also been shown that bond lengths in blue and red PDA do not differ significantly. It was proposed early on that red chains are non planar, successive repeat unit being rotated in opposite directions relative to the chain's average plane \cite{Tanaka89}, so that the 1D unit cell contains two non translationally equivalent repeat units; we believe that this is the most likely explanation despite the fact that several red PDA crystal structures are not reported with a doubled unit cell \cite{Kobayashi87, Foley99}. A possible explanation for this discrepancy has been proposed in \cite{Schott06b}. A recent DFT theoretical calculation \cite{Filhol09} indeed relates a stable non planar chain conformation to the spectroscopic properties (visible absorption, vibrational frequencies, NMR chemical shifts) in a simple model PDA and in the known PDA poly-THD \cite{Enkelmann80, Morrow87, Agrinskaya95, Barisien07}.\\
The presence of low energy dark states has been related theoretically to the importance of electronic correlations, as was claimed before for polyenes \cite{Ovchinnikov73, Tavan86}. However, theoretical calculations on PDA generally fail to find low enough energies for the dark states \cite{Race01, Race03}. These calculations dealt exclusively with planar PDA chains.\\
To understand the relationship between conformation and electronic structure in PDA, a better knowledge of electronic properties of non planar chains is needed, also providing new experimental data to be compared with further theoretical studies. Obtaining high quality isolated chains dispersed in their monomer single crystal is less demanding than obtaining a perfect completely polymerized PDA crystal: for instance poly-4BCMU and poly-3BCMU chains in their monomer crystal are high quality quantum wires, whereas the 4BCMU crystal becomes disordered around a polymer concentration of 10$\%$ in weight \cite{Brouty88, Spagnoli07a} and 3BCMU crystals do not even polymerize completely \cite{Patel81}. Yet, absorption spectra of chains in crystals with low polymer content  are often not well resolved. Therefore, we have engaged in a search of reactive diacetylenes (DA) forming high quality single crystals that would contain isolated red chains only. \\
In a previous paper \cite{Barisien07} we reported on poly-THD, the crystal structure of which is known; its unit cell contains two non equivalent repeat units, tilted by $\pm$ 15$^{\circ}$ from the average plane of the chain \cite{Enkelmann80}. Poly-THD isolated chains are strongly fluorescent at low temperature \cite{Agrinskaya95, Barisien07}, but the fluorescence yield rapidly decreases with an increase in T, apparently due to the opening of an efficient non radiative decay channel for the exciton, quite similar to the situation in poly-3BCMU red chains \cite{Lecuiller02}. The exciton absorption peaks at 571 nm (or 2.17 eV) at 15 K, intermediate between the usual values of blue and red chains. Recent calculations of the poly-THD crystal ground state structure \cite{Filhol09} have found three energy minima, one of which being identical to the observed structure. Indeed, for a given set of bond lengths and angles, there is only one possible planar chain structure, while different tilt angles may occur in non planar chains. This suggests that it might be possible to obtain a set of PDAs with different tilt geometries making possible the study of the effects of electronic correlations as a function, for instance, of chain geometry.\\
In the present paper, isolated chains dispersed in their monomer crystal matrix of a newly synthesized DA \cite{Deschamps} are studied. The monomer crystal structure shows that, as in THD, the 1D unit cell of the chain contains two repeat units and suggests that the chains will be non planar (the exact conformation of isolated chains at small concentration is not experimentally accessible yet). It is shown that the exciton energy is higher than that in usual red PDA. As a result, one now has a series of four types of PDA chains (blue poly-3BCMU or poly-4BCMU, poly-THD, red poly-3BCMU and the present material) showing a regular change in energies and relative positions of the excited states.\\
Moreover, it will be shown that the chains in this material have the optical properties of highly luminescent quasi-1D systems, as isolated red poly-3BCMU chains do \cite{Lecuiller02, Dubin06a, Dubin06b}. But in the DA studied in this paper these properties are shared by all chains in the crystal and are not restricted to a small minority of chains, so studies that were not possible on poly-3BCMU such as electroabsorption as in the attendant paper \cite{Weiser}, non linear absorption or pump-probe spectroscopies become feasible.
The spectroscopic properties of these chains will be described and discussed in two companion papers, this one and another \cite{Weiser} essentially dealing with the study of electroabsorption. The present paper deals with absorption and luminescence spectra, luminescence temporal decay, all as a function of temperature. 
\section{Materials and experimental methods}
The formula of the studied DA is (C$_{6}$H$_{5}$)$_{2}$N-(CH$_{2}$)$_{3}$-C$\equiv$C-C$\equiv$C-(CH$_{2}$)$_{3}$-N(C$_{6}$H$_{5}$)$_{2}$. Its synthesis is described in \cite{Deschamps}. We propose to name it 3NPh2, 3 being the number of CH$_{2}$ groups and NPh2 describing the rest of the side group. This method has been used to name DA of the BCMU family for instance. This molecule is an analog of THD, that contains a single CH$_{2}$ group and that could be named 1NPh2 according to the aforementioned naming scheme.

\subsection{Monomer crystal structure and consequences for reactivity}
\begin{figure}
\includegraphics[scale=0.8, width=7cm, height=6cm]{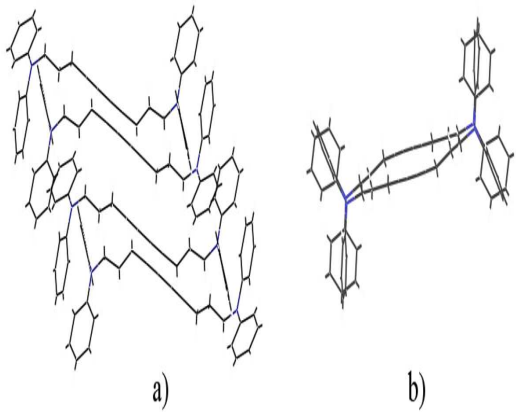} 
\caption{Crystal structure of the 3NPh2 diacetylene: a) Two pairs of non equivalent monomers; b)The same pairs as in a) but viewed along the polymerization direction (\textbf{a} axis), perpendicular to the plane of the figure.}
\label{Figure1}
\end{figure}

High quality large platelike single crystals, with typical dimensions 5$\times$1$\times$0.4 mm were grown from acetone solutions at 4$^{\circ}$C in the dark and stored at –18$^{\circ}$C. They slowly polymerize under $\gamma$ irradiation so the polymer content, x$_{p}$, can easily be adjusted at will. Since the monomer is not thermally reactive, the polymer contents stay constant after irradiation.\\
3NPh2 crystallizes in space group P21/n with Z = 4. At 187 K \textit{a} = 9.9484(7), \textit{b} = 16.0893(12), \textit{c} = 16.7643(14) all in \AA, $\beta$ = 98.697(4) degrees and V = 2652.5 \AA$^{3}$. No phase transition has been detected via optical measurements between 4 and 300 K. The structure is further analyzed in \cite{Deschamps}.\\
Polymer chains grow parallel to the \textbf{a} axis, so the average repeat distance between reactive C4 groups is 4.974 \AA ° at 180 K, but this average value is misleading since there are along \textbf{a} two types of non equivalent monomers, with differently oriented terminal phenyl rings (and other less important geometrical differences). These monomers are arranged in pairs, with the intermonomer distance along \textbf{a} within a pair smaller than the distance between monomers belonging to different pairs. Two consecutive pairs are not coplanar, but are alternatively on either side of the \textbf{a} axis. The packing of successive monomers viewed along the polymerization direction is sketched in figure \ref{Figure1}b. 

The geometry of a ``close'' pair is favorable to chain initiation: the distance between the two C atoms which will form a double bond in initiation is 3.49 \AA ° and the intermonomer distance along \textbf{a} is 4.93 \AA ° at 180 K. These values are not far from the optimal ones \cite{Baughman74, Baughman78, Enkelmann84}. However, one should bear in mind that polymerization is performed at 300 K. A reasonable estimate of the expansion coefficient along \textbf{a} is $\alpha_{a}$ $\approx$ 9$\pm$1 10$^{-5}$K$^{-1}$, a typical value for van der Waals crystals, so the room temperature intermonomer distance would be $\approx$ 4.98 \AA, a slightly less favorable value.\\ 
But the distance between two neighboring monomers not belonging to the same pair is much larger, 5.39 \AA ° at 180 K, that is $\approx$ 5.43 \AA ° at 300 K, outside the range where reaction is observed \cite{Enkelmann84}; there is no van der Waals contact between them.\\
This geometry suggests that initiation within a pair may be relatively easy, subject to conditions on side group displacements which must avoid interatomic penetration, while formation of a dimer from two neighboring monomers not belonging to the same pair will not occur even though the distance between the C atoms that would bind by initiation is only 3.57 \AA. Propagation of course must involve interpair bonding, but this implies an active chain end and a neutral monomer, not two monomers as initiation does. Careful studies of photopolymerization of the DA pTS \cite{Sixl84} have shown that the active chain end is a carbene except for the first few propagation steps where it is a radical. The C4 axis of the active end is rotated towards the chain growth direction with which it makes an angle $\approx$ 29$^\circ$, as compared to 45$^\circ$ in the starting monomer and 14$^\circ$ in the final polymer chain \cite{Bubeck80, Hartl82}. So, topochemical conditions for propagation are generally less demanding than those for initiation.\\ 
It is reasonable to assume that this initiation and propagation scheme also holds true for 3NPh2, explaining the occurrence of chain propagation leading to the observed long chains. However, we are dealing here with $\gamma$-induced polymerization and it has been shown for the DA 3BCMU and 4BCMU that ionic initiation must play a role, possibly dominant \cite{Spagnoli07b}. But whatever the exact electronic structure of the active end, it must have rotated towards the \textbf{a} axis since all initiation processes lead to formation of the same PDA chain.\\
\subsection{Polymer chain geometry}
Of course the details of the geometry will be affected by chain formation, for instance it is likely that the C4 groups will move closer to the \textbf{a} axis. But the side group will still force a double unit cell and the space occupied by neighbouring monomers does not allow that all repeat units of the chain become translationally equivalent: a non planar chain is expected, with two non equivalent repeat units by 1D unit cell.
Polymer chains formed in the early stages of polymerization, such as those studied in this paper, are embedded in a mostly monomer crystal. If the chain stays in register with the crystal as has been experimentally demonstrated in the DA pTS and DNP \cite{Albouy82, Albouy85}, then the 1D polymer unit cell length will be equal to the \textit{a} parameter. To estimate \textit{a} at low temperature where most of the present results have been obtained, we assume for simplicity that the same value of the thermal expansion coefficient, $\alpha_{a}$, applies down to 80 K, and that it becomes 0 at lower T. These rough assumptions yield \textit{a} $\approx$ 9.75 \AA. The equilibrium repeat distance of almost all known bulk PDA crystals is 4.90 $\pm$ 0.02 \AA \cite{Enkelmann84, Schott87} so in 3NPh2 the relaxed polymer unit cell is expected to be $\approx$ 9.8 \AA.\\ 
Therefore isolated poly-3NPh2 chains are expected to be longitudinally unstrained or nearly so at low T. They become under increasing tension as T increases, up to a strain $\approx$2.2 $\%$ at room T. Comparable tensile strains on bulk PDA crystals such as poly-DCH  and poly-pTS do not cause major changes in the electronic properties (no ``color transition'') \cite{Batchelder78}, and similar compressive strains on isolated poly-4BCMU chains \cite{Spagnoli07c} do not either.
\subsection{Crystal optics}
The large surface of the platelike crystals is the (\textbf{a,c}) plane. Therefore chains grow parallel to the surface and the incident light can be polarized parallel or perpendicular to the chain direction. But the binary axis \textbf{b} is perpendicular to it, so in optical experiments light propagates parallel to the binary axis, and its polarization is in the (\textbf{a,c}) plane, so always perpendicular to the binary axis. A similar situation holds in some DA such as DNP, which belongs to the same space group as 3NPh2 \cite{MacGhie81, Helberg94}. This differs from the most common situation in monoclinic DA (pTS or 3BCMU for instance) where the binary axis is also the chain direction and lies in the surface plane, so that light propagates perpendicular to the binary axis and its polarization can be made parallel or perpendicular to it.
Monoclinic crystals are optically biaxial. In addition weakly polymerized crystals are absorbing media and the directions of the principal axis of the [$\epsilon_{1}$] and [$\epsilon_{2}$] tensors of the real and imaginary parts of the dielectric constant will not be the same, in general. Given the low x$_{p}$ values of the studied crystals, the optical properties can be investigated in the weakly absorbing limit. In this situation the following simplified statement applies \cite{Born_Wolf, Beugnies}: two normal modes can propagate with their \textbf{k} vector parallel to the binary axis \textbf{b} while conserving their linear polarization along the two optical axes of [$\epsilon_{1}$] (perpendicular to \textbf{b}). Their amplitudes are damped differently during the propagation, depending on the angle between the optical axis and the direction defined by the chains (parallel to \textbf{a}). In our crystals, experiment clearly shows that there is a weakly absorbed mode and a strongly absorbed one with a high dichroïc ratio. One of the axes of [$\epsilon_{1}$] is very close to the \textbf{a} direction (see inset of figure \ref{Figure5}) which is certainly one axis of [$\epsilon_{2}$] since the chains are the sole absorbers in the visible range. Observation of the crystal between crossed polarizers at ambient temperature indeed allows to determine the angle $\delta$ between the optical principal axis and the crystal edge (namely the \textbf{a} direction). We find $\delta  \sim 2.5 ^\circ$ with a $\pm$1$^\circ$ accuracy. However a modification of this geometry may occur with varying temperature. In absorption experiments the reference direction will be the most absorbing direction (field \textbf{E // a}). In the experiment reported below a second polarizer is placed after the sample, in order to select one of the polarization directions of [$\epsilon_{2}$].
\subsection{Polymerization}
3NPh2 monomer crystals placed in glass tubes were irradiated in air in a $^{137}$Cs $\gamma$-ray source (IBL 637 – CIS BIO International – number 95-22 at the Research Section of the Curie Institute in Paris) at a dose rate of $\sim$ 40 Gy/min for durations between 30 s and 20 min. Crystals turn bright yellow upon polymerization, a colour that differs from those of other PDA crystals known to date.
Unlike poly-THD, poly-3NPh2 is soluble in several common organic solvents (for instance THF, chloroform), allowing determination of the absorption coefficient of the polymer chains in the monomer crystal \cite{Deschamps}. Hence, the polymer content x$_{p}$ is accessible by the method described in \cite{Spagnoli96}.
Pure poly-3NPh2 was obtained as follows: the irradiated crystals were placed in acetone to dissolve the unreacted monomer. The insoluble material (poly-3NPh2) was collected by filtration, washed several times with acetone, and air-dried. The pure polymer in good solvent has the same absorption spectrum as other PDA in the same conditions, a single broad symmetric band peaking at 470 nm. This indicates that poly-3NPh2 also forms semi-rigid chains with a persistence length of about 150 \AA \cite{Wenz84, Rawiso88}.
\subsection{Experimental methods}
Transmission spectra were measured in the double beam mode of a Cary 5000 double spectrometer. The crystals were placed on thin nickel optical masks perforated with rectangular slits (width 100 $\mu$m, length 2 mm). For temperature dependent studies, the slit was fixed on the cold finger of a commercial helium exchange gas cryostat. The temperature was monitored between 13 K and ambient temperature by Si diodes and controlled by heating of the gas. A second identical slit was centred through the reference beam to balance the setup. The baseline was repeatedly checked as it slightly varies with temperature. A spectral resolution of 0.4 nm was generally sufficient except when rapidly varying signals were expected, in which case 0.2 nm was used. Anisotropic properties of the absorption were investigated by inserting polarizing foils in the probe beam. One foil sets the incident polarization whereas the other one placed after the sample is used to analyze the transmitted light. For compensation purposes the reference beam was equipped with identical polarizers along the probe line.
A confocal-like microscope configuration was used to investigate the luminescence properties between 2.4 K and 80 K. The excitation beam was focused using a microscope objective with NA=0.6, yielding a waist size of $\sim$1 $\mu$m. Luminescence was collected with the same objective, an additional stage consisting of a pair of lenses and a pinhole (diameter 50 $\mu$m) centred on the optical path was used as spatial filter. The luminescence was analyzed in a Triax-550 Jobin-Yvon spectrometer coupled to an N$_{2}$ cooled CCD camera with an overall spectral resolution of $\sim$ 100 $\mu$eV. In the experiments, the sample was mounted on the cold finger of a micro-photoluminescence cryostat (Oxford Instruments) designed to compensate for thermal expansion and keep a fixed spatial position of the spot as temperature is changed. Time resolved luminescence experiments were performed by coupling a streak camera (Hamamatsu C5680) to the second exit of the spectrometer.
\section{Optical absorption}
\subsection{Unpolymerized monomer crystal}
As-prepared monomer crystals, before any polymerization by irradiation or heating, are very slightly absorbing in the polymer region where the monomer is transparent. This is almost always observed in reactive DA crystals. The origin of these polymer chains is unknown, but they probably correspond, either to absorption of background radiation (cosmic rays, natural radioactivity of pyrex glass used in crystal growth etc...) or to thermally induced initiation at special defect states with low activation energy.
A typical absorption spectrum at low T, $\alpha_{//}$, for light polarized parallel to the \textbf{a} axis, the direction of chain growth, is shown in Figure 2. It consists of an origin line near 2.4 eV and several broad bands near 2.55, 2.65 and 2.77 eV. Such a spectrum is typical of the exciton absorption of polydiacetylenes with a zero-phonon exciton band and its associated vibronic satellites.
\begin{figure}
\includegraphics[scale=0.45, width=8cm, height=7cm]{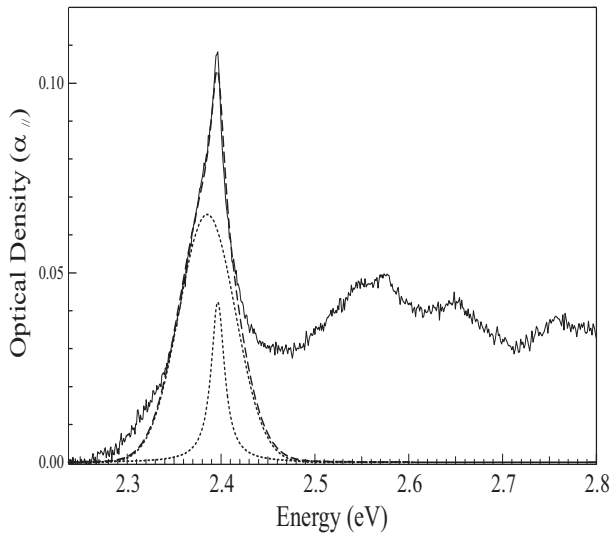} 
\caption{Absorption spectrum, at 13 K, of a non irradiated crystal (solid line). The main exciton line is the sum of a Lorentzian contribution ($\Gamma \sim $17 meV) and a Gaussian contribution ($\Gamma \sim $68 meV, dotted line). Result of the fit: dashed line.}
\label{Figure2}
\end{figure}
In figure \ref{Figure2}, the exciton absorption band has a somewhat peculiar lineshape, consisting of a fairly broad band at 2.387 eV with a superimposed much narrower component peaking at 2.397 eV. A fit using such two components is shown in the figure. The fitted region can be extended to higher energies by inclusion of broad bands corresponding to vibronic satellites. It is noteworthy that the narrow component is less conspicuous in the vibronic satellites, suggesting a larger Frank-Condon factor for it.
As will be shown below, the narrow line corresponds to chains identical to those prepared by $\gamma$ irradiation. Using the corresponding absorption coefficient determined below, their concentration is $\sim$10$^{-6}$ in weight in this particular sample. The area of the broader band is approximately 6 times larger so, assuming the same oscillator strength, the concentration is $\sim$ 6$\times$10$^{-6}$ for this different type of chains. Thus before irradiation, the crystals contain a very small polymer concentration, on the order of 10$^{-5}$ or less.
The broad Gaussian component does not increase significantly upon $\gamma$-irradiation and, as x$_{p}$ increases, it becomes a negligible contribution to the overall absorption response, indicating that the amount of specific sites it originates from has been exhausted. The following fits will however systematically take it into account as a small correction, nevertheless important for a precise analysis of the excitonic line structure.
\subsection{Overall absorption spectrum and its temperature dependence}
\begin{figure}
\includegraphics[scale=0.45, width=8cm, height=7cm]{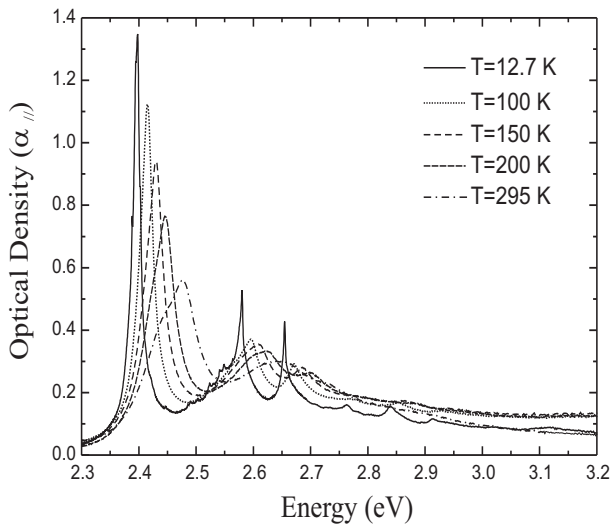} 
\caption{Temperature dependence of the optical absorption, $\alpha_{//}$ (incident electric field parallel to the \textbf{a}-axis), of isolated poly-3NPh2 chains in their host crystal (x$_{p} \sim$ 4$\times$10$^{-5}$). The sample thickness e$\sim$ 440 $\mu$m leads to an absorption constant at the signal maximum, $\alpha_{max} \sim $ 72 cm$^{-1}$ at 12.7 K.}
\label{Figure3}
\end{figure}

Figure \ref{Figure3} shows spectra recorded at various temperatures after very short $\gamma$ irradiation. The polymer content was $\sim$4$\times$10$^{-5}$ in weight. This spectrum is therefore expected to correspond to truly isolated chains. The lowest temperature spectrum can be compared to figure \ref{Figure2}. It is now dominated by the ``narrow component''. Again this is a typical PDA excitonic spectrum. At 13 K, the pure exciton line is centered at E$_{0} \sim$ 2.399 (6) eV ($\lambda_{0}$ = 516.7 nm) which, to our knowledge, makes poly-3NPh2 resonance energy the highest reported among well ordered PDA. The measured exciton linewidth (FWHM) is $\sim$ 17 meV. It is slightly larger than for isolated chains in other PDA, but this is still a very low value, indicating that we are dealing with high quality quasi-1D isolated chains \cite{Schott06a}. A long vibronic progression characteristic of the backbone develops on the high energy side. The so-called D and T lines associated to the C=C and C$\equiv$C normal stretching modes of the chain are clearly identified at E$_{D}$=2.582 (5) eV and E$_{T}$=2.657 (2) eV as well as their harmonics and combinations at 2.763 (8) eV (2D), 2.838 (5) eV (D+T) and 2.913 (2) eV (2T). The D and T bands show up as very narrow peaks, and a number of weaker narrow peaks appear on the low energy side of the D line, above 2.5 eV. In addition, a weak broad band is present at $\sim$ 3.12 eV, which does not belong to these vibronic progressions. It is shown in the companion paper \cite{Weiser} that this structure corresponds to a weakly bound exciton close to the dissociation threshold, which can be observed here because the larger exciton binding energy displaces farther from the gap energy the overall vibronic progression of the lower energy exciton.
Between ambient temperature and 13 K all absorption lines are red-shifted by $\sim$ 80 meV. The smooth variation indicates that no significant structural modification occurs as temperature is decreased.
The dichroism in absorption, $\rho = \alpha_{//}/\alpha_{\perp}$ was determined at room temperature on crystals having higher x$_{p}$ ( x$_{p} \sim $10$^{-4}$) so that $\alpha_{//}$ and $\alpha_{\perp}$ could be measured on the same sample. $\rho$ was found to be equal to 220 $\pm$ 20. This large value confirms the high level of order of the monomer crystal and compares favorably with values found in 4BCMU and THD \cite{Spagnoli94, Barisien07}. It is worth noting that, in the present situation, $\rho$ is measured by placing an analyzer behind the sample along the optical path. As a consequence, $\rho$ leads to a ratio which compares the strength in absorption between two directions given by the [$\epsilon_{2}$] principal axis. Due to the possible misorientation existing between the chains and the principal optical axis (of [$\epsilon_{1}$]) it is clear that $\rho$ does not measure a ratio which compares the strength of the excitonic dipolar transition in its own direction and that in the perpendicular one. $\alpha_{\perp}$ overestimates the actual intrinsic absorption of the chains in the perpendicular configuration; we thus expect an actual intrinsic dichroism larger than the presently measured dichroism.
\subsection{Influence of birefringence}
\begin{figure}
\includegraphics[scale=0.45, width=8cm, height=7cm]{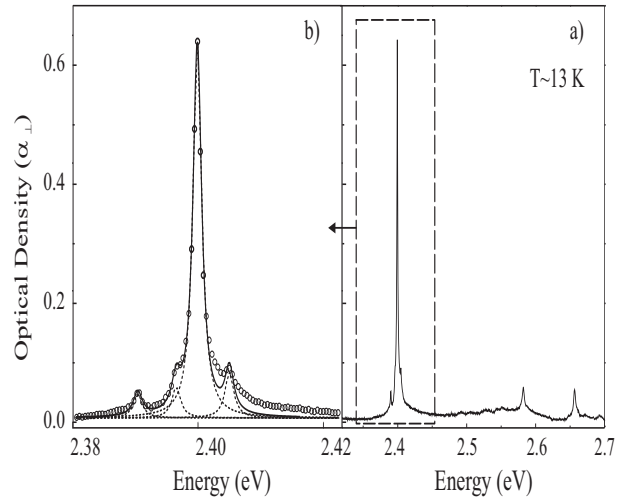} 
\caption{a) Absorption spectrum, $\alpha_{\perp}$, at 13 K of a crystal with x$_{p} \sim $8$\times$10$^{-4}$ for an incident field polarized perpendicular to the chain direction. b) Detail of the exciton line: experimental data (open circles), fit (solid line) and lorentzian contributions to the fit (dashes).}
\label{Figure4}
\end{figure}

The shape of the exciton line is complex: it always shows evidence of several narrow components. In perpendicular polarization (figure \ref{Figure4}) four components are clearly resolved: the dominant contribution peaks at E$_{0} \sim $ 2.399 (6) eV with weaker lines at 2.390 (5), 2.396 (4) and 2.404 (7) eV (at 13 K).

\begin{figure}
\includegraphics[scale=0.45, width=8cm, height=7cm]{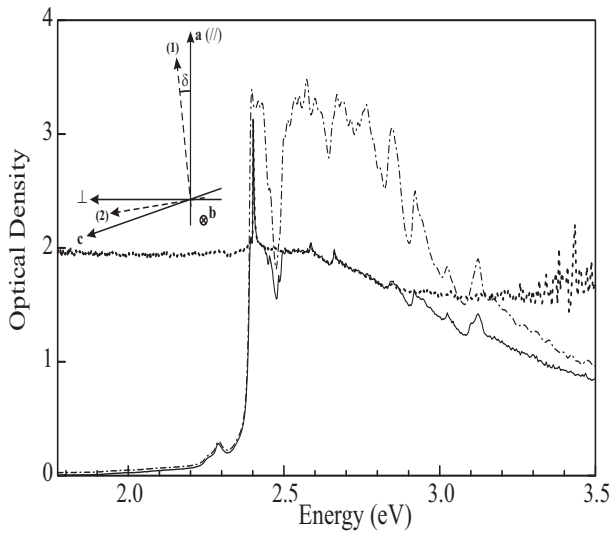} 
\caption{Absorption spectra, at 13 K, of a crystal with x$_{p} \sim $1.4 $\times$10$^{-3}$ for different configurations of the polarizer and analyzer. Dash-dotted line: //-// configuration; solid line: //-(no analyzer) configuration; dots: //-$\perp$ configuration. Inset: geometrical configuration for absorption spectroscopy of 3NPh2 crystals. \textbf{a, b, c} are the crystallographic axis. (1) and (2) are two of the principal axis of [$\epsilon_{1}$]; \textbf{b} is the third principal direction in a monoclinic crystal. $\delta$ is the angle between \textbf{a} and (1). In the experiments the incident light is perpendicular to the (a,c) plane and polarized either parallel (//) or perpendicular ($\perp$) to \textbf{a}.}
\label{Figure5}
\end{figure}

Those components also show up in the $\alpha_{//}$ spectrum but are more clearly resolved in $\alpha_{\perp}$ due to the extreme narrowness of the lines. The width, $\Gamma_{\perp} \sim$ 1.8 meV, is the same for each line. It is significantly smaller than the exciton width extracted in the // spectrum ($\Gamma_{//}$ / $\Gamma_{\perp} \sim $3) in which lines at the same energies are observed. This large difference indicates excitons with different scattering rates namely different degrees of coupling with the surrounding medium according to the nature of the excited dipole; this has not been investigated further in the present situation however comparable observations have already been made. For instance the absorption spectra of high quality Naphtalene crystals show differences of one order of magnitude in linewidths for the weakly and strongly absorbing Davydov component of the same excitonic transition \cite{Robi78}. This is also observed for the fundamental exciton of $\alpha$-sexithiophene. The broader linewidth of the strong Davydov component is explained by a more rapid migration of these excitons, due to their larger dipolar coupling to excitons of neighboring cells, which allows them to visit more defects in a given time \cite{Moller2000}.
As shown below, the higher energy weak line at 2.404 (7) eV seems to be the origin of a particular luminescence spectrum, while the other weak lines do not show up as different origins in the overall luminescence. Observation of several exciton lines is fairly common in PDA chains diluted in their monomer crystal matrix \cite{Schott06a}, but their study in the present material with the previously used methods would be technically difficult. This is further discussed below.\\
The influence of optical birefringence onto the absorption properties was evidenced in slightly more polymerized crystals. Figure \ref{Figure5} displays three absorption spectra corresponding to different configurations of the polarizers and analyzers in a sample with x$_{p} \sim $ 1.4$\times$10$^{-3}$ namely $\sim$ 1.7 times more polymerized than the sample which response in perpendicular polarization is shown in figure \ref{Figure4}.

As expected from the larger polymer content, the spectrum has a characteristic saturated shape when the incident light is polarized parallel to the \textbf{a} axis and analyzed in the same direction. Between $\sim$ 2.37 eV and 2.75 eV the saturation value of the absorption is exclusively determined by residual stray light (O.D.$\sim$3.3). At lower x$_{p}$, this configuration would measure, in good approximation, the absorption coefficient associated to the highly damped mode (mode (1) as shown in the inset of figure \ref{Figure5}).
When the analyzer is removed a truncation effect is still visible but in comparison with the previous situation extra sharp peaks appear on top of the plateau which is associated to an appreciably reduced optical density around 2. When subtracting the plateau contribution the spectrum becomes identical to the optical response measured in perpendicular polarization, the main vibronic replica associated to the C=C and C$\equiv$C stretching modes being perfectly visible. The lower O.D. indicates that, except in some regions where resonances of the chains can be distinguished, more light is transmitted by the sample. This light is the contribution to the transmittance of the weakly absorbing mode (2) which can be measured separately, with a proportionality factor cos$^{2}(\delta)$, by placing an analyzer perpendicular to the \textbf{a} axis, the polarizer on the incident beam being kept parallel to the a direction (see figure \ref{Figure5}).
In the latter configuration a plateau is also observed in the transparency region and gives in principle, access to the angle $\delta$. Below 2.0 eV, each mode indeed contributes equally to the transmittance so that the O.D. simply relates to $\delta$ according to:
\begin{eqnarray}
10^{-O.D.} \sim 2\times cos^{2}(\delta)sin^{2}(\delta) \sim 2sin^{2}(\delta)
\end{eqnarray}

which leads, for O.D.$\sim$2,  to $\delta \sim 4 ^\circ$ in good agreement with the ambient temperature measurement given that a variation of $\delta$ with temperature cannot be excluded as well as an additional misorientation originating from the sample positioning. A more quantitative estimation of $\delta$ through absorption measurements was not attempted in the present work.
\section{Luminescence}

\begin{figure}
\includegraphics[scale=0.45, width=8cm, height=7cm]{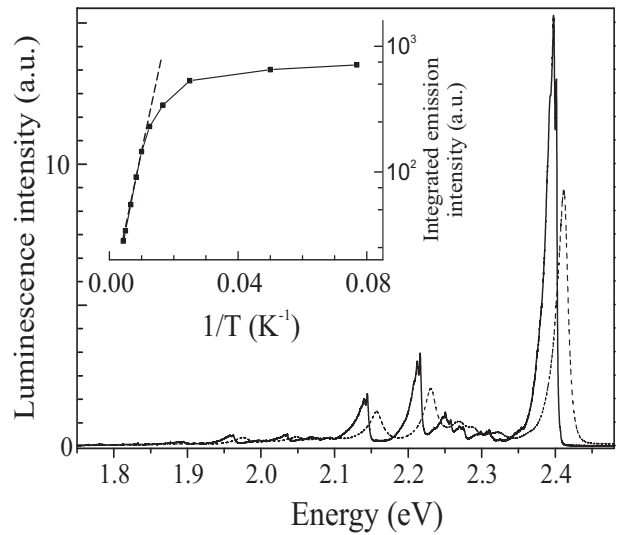} 
\caption{: Emission of isolated chains of poly-3NPh2 at 8 K (solid line) and 55 K (dashed line). Inset: Arrhenius-like plot of the integrated emission intensity (solid line) and associated fit for T $\succ$ 60 K (dashes) giving an activation energy E$_{act} \sim $22 meV.}
\label{Figure6}
\end{figure}

The luminescence of poly-3NPh2 is intense enough to be easily detected up to room temperature. Figure \ref{Figure6} displays spectra measured at 8 and 290 K on a sample with very low x$_{p}$ ($\sim$ 3$\times$10$^{-5}$ weight) under resonant excitation of the T line. Polarization of the excitation was set parallel to the \textbf{a} axis.
At 8 K the luminescence spectrum is almost a mirror image of the absorption signal. Numerous vibronic satellites can be identified, each one having its own counterpart in the absorption spectrum with, below 30 K, a fine structure on the main vibronic lines which is associated to a low energy phonon and will not be discussed here. The luminescence follows the absorption shift with temperature between 8 K and 290 K. Its dichroic ratio was measured as $\rho_{L} \sim $120 at 8 K and found to be independent of the wavelength. The narrow and broad emissions also respond similarly to the incident beam polarization.

\begin{figure}
\includegraphics[scale=0.45, width=8cm, height=7cm]{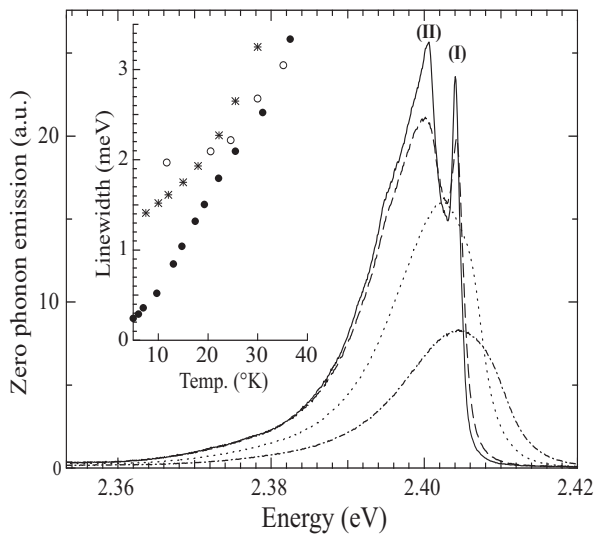} 
\caption{Zero phonon emission of poly-3NPh2 chains at 8 K (solid line), 18 K (dashed line), 30 K (dotted line) and 60 K (dash-dotted line). Inset: Temperature dependence of the linewidth of different polydiacetylene chains: isolated poly-3NPh2 in crystals with x$_{p} \sim $3 $\times$10$^{-5}$ in weight (stars), ensemble of isolated poly-3BCMU chains x$_{p} \sim $3 $\times$10$^{-5}$(open circles) and single poly-3BCMU chains at x$_{p} \sim $3 $\times$10$^{-7}$ (filled circles).}
\label{Figure7}
\end{figure}

Figure \ref{Figure7} shows the zero phonon emission at high resolution and on an extended scale. It consists of a very narrow well resolved peak with a Lorentzian profile at 2.404 eV (referred to as emission (I) henceforth) and an asymmetrically broadened peak (emission (II)), peaking at 2.400 eV. Both coincide with transitions in the absorption. All vibronic lines in emission also present a narrow component which corresponds to a vibron of the zero phonon emission (I). 
Since the linewidths rapidly increase with T the narrow peak is visible only up to 40 K. The change as a function of temperature of the emission linewidth of (I) is plotted in the inset of Figure 7. The lorentzian shape on the high energy side holds up to 30 K and is used as reference for the fit. For comparison, data related to isolated poly-3BCMU wires are also shown \cite{Dubin02}. Single poly-3BCMU chains have Lorentzian lineshapes at all temperatures, while the width of an ensemble of such chains shows at the lowest temperature some inhomogeneous broadening with a saturated value of $\sim$2 meV, but the homogeneous width is recovered above $\sim$40 K. Here we are dealing with such an ensemble of chains. Their Lorentzian widths are slightly larger than in poly-3BCMU, corresponding to a slightly shorter dephasing time, but inhomogeneous broadening is much less important; whether the slower decrease of $\Gamma$ with temperature below 20 K arises from a change in dephasing processes, or to incipient effects of inhomogeneity is not clear, but the inhomogeneous width is certainly less than 1 meV.

The inset in figure \ref{Figure6} shows the temperature evolution of the spectrally integrated luminescence. The contribution of the narrow emission remains very small at any temperature and the following analysis applies only to the broader part of the emission. Above 60 K, the intensity variation is compatible with a thermally activated behaviour of the non radiative rate, k$_{nr}$, associated to the excitonic level. Assuming that k$_{nr}$ becomes much larger than the radiative rate k$_{r}$ in this temperature range, we find here E$_{act} \sim $22 meV with a rather small preexponential factor, k$_{nr,0}$ (we estimate k$_{nr,0}$ $\sim$ 4$\times$10$^{11}$ s$^{-1}$). Hence, a non radiative decay channel opens at an energy just above the energy of the \textbf{k}=0 state of the exciton band. Such a behaviour has been observed in other red PDA. For instance in poly-3BCMU chains, k$_{nr}$, is constant at low T and starts increasing above 50 K at a rate compatible with an activation energy $\sim$ 30-40 meV \cite{Lecuiller02}. The nature of that channel is not studied here. One may assume that there is an energy level just above that of the exciton. However, another process has been identified for poly-4BCMU and poly-3BCMU blue chains: the yield of formation of triplets from singlet excitons is very small at the exciton energy but increases quickly at higher energy to reach a plateau $\leq$ 200 meV above it, and this has been shown to correspond to the onset of singlet exciton fission S$\rightarrow$2T \cite{Kraabel98b}. In this process, the threshold is determined by the energy of a triplet pair, and does not correspond to a single excitation state. A pump-probe study of triplet-triplet absorption could answer this question.

\begin{figure}
\includegraphics[scale=0.45]{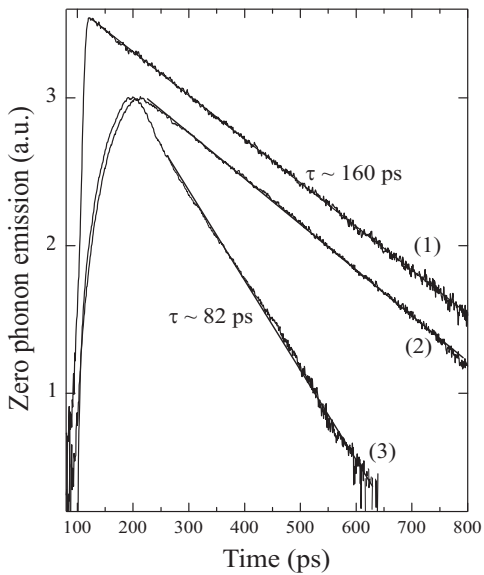} 
\caption{Temporal decay at 6.5 K of the poly-3NPh2 zero phonon emission. (1) and (2): Zero phonon ``broad" emission (II); (3) dynamics of the ``narrow" emission (I). Curve (1) is obtained using a twelve times less dispersive grating (100 lines/mm). The pump power is always less than 50 nW.}
\label{Figure8}
\end{figure}

Time resolved experiments with picosecond resolution were also performed. The excitation power was kept below 50 nW in order to avoid heating. At any temperature the decays are well fitted by a single exponential over more than two orders of magnitude (Figure 8). Each type of emission has its own dynamic. An effective lifetime, $\tau_{eff}$, of 160 ps is measured for the broad emission at 8 K whereas the decay associated to emission (I) is about twice as short.
One should note that the emission spectrum does not change in shape during the decay; a similar dichotomy in dynamics is measured in each vibronic band with the same characteristic constants. It is also important to point out the negligible influence of re-absorption in the zero phonon region as far as dynamics is concerned. Identical lifetimes are indeed deduced from the fits when comparing the zero phonon and vibronic emissions, the latter not being subject to re-absorption. This is explained on one hand, by the low x$_{p}$ of the probed samples and on the other hand, by the experimental reflection geometry chosen for the experiment. The confocal-like configuration allows the selection of light coming from a thin layer under the outer surface. The longest path, in the crystal, of detected luminescence is then typically on the order of one micron.
Note that, in figure \ref{Figure8} the experimental rise time of the emission signal is determined by the instrumental response. It is considerably lengthened when more dispersive gratings are used (compare curve (1) with curve (2)). It was thus carefully checked that the decays were not affected when high spectral resolution was required especially for the simultaneous study of both components.
\section{Discussion}
\subsection{Exciton absorption lineshape}
The observation of weak extra lines in the exciton region is usual in polydiacetylene crystals provided that the widths are small. In general the separation is relatively large, $\sim$ 20 meV or more and their general properties are similar to the main excitonic ones. In 3NPh2 crystal the situation differs by at least two points: the absorption line at 2.404 (7) eV is associated to a strong resonant luminescence with different $\tau_{eff}$ and the energy separation between the satellite lines and the zero-phonon line is at maximum $\sim$10 meV.
The different lines could correspond to the response of excitons on chains with very slightly different conformations and therefore slightly modified electronic structures. This implies the existence of minority chains populations coexisting in the crystal with the dominant population absorbing at E$_{0} \sim$ 2.400 eV. Observation of a fluorescence resonant with the absorption transition at 2.404 eV demonstrates that this transition indeed is that of a different chain population. The generation of those chains in the vicinity of specific sites of the monomer crystal is a possible explanation. Similar weak lines have also been observed in other DA-PDA mixed crystals at low polymer contents, namely in partially polymerized 3BCMU and 4BCMU crystals. In these materials, it is clear that the weak absorption lines are the excitonic transitions of minority populations of chains, differing slightly from the main one: Raman frequencies are slightly different, showing that these chains have slightly different ground state geometries \cite{Spagnoli94, Lapersonne94}. In addition, the photo-bleaching of a given weak transition does not bleach the main absorption line so that the lines have to be assigned to independent chains \cite{Kraabel98a}. It was moreover demonstrated in electro-absorption experiments that there is no transfer of oscillator strength between the different transitions \cite{Horvath96}. 
An alternative explanation for the satellites structures in absorption could be the coupling of the exciton with a low energy vibration mode. However the energy separations are irregular, so this possibility would require a specific mode for each satellite line. This is a highly unlikely situation.
Note that in the present case the small energy separation makes difficult the application of the methods previously employed to address the different populations, such as a selective excitation in Raman or pump probe spectroscopy. As a consequence, the study was not further continued in that direction.

\subsection{Temperature shift of transitions}
In DA-PDA mixed crystals produced during the topochemical polymerization, the shift in polymer exciton energy with T may be due (at least) to changes in polarization of the surrounding medium or to changes in the chain electronic structure itself, corresponding to a modification of the geometry of the chain.
In van der Waals solids, an electronic state energy contains a term D due to electronic polarization of the solid medium which tends to decrease the energies \cite{Craig68, Schott87}. Generally, the excited states polarizabilities are larger than that of the ground state; so if $\mid$D$\mid$ increases, the energy of excited states decreases faster than that of the ground state and the transition energy decreases. Decreasing T means decreasing volume hence increasing the electron density of the medium, hence of $\mid$D$\mid$.
But decreasing T may also mean decreasing the repeat unit distance of the polymer chains in the mixed crystal, hence a change in the chain geometry.
The two processes can be seen at play in the blue chains of poly-3BCMU and poly-4BCMU \cite{Schott06a}. In the former there is almost no change in repeat unit distance between 300 and 15 K \cite{Spagnoli07a}so the change in $\mid$D$\mid$ dominates. The resulting decrease in exciton transition energy is -50 meV. In poly-4BCMU the repeat unit distance decreases by about 2.5$\%$ in the same T range and the exciton energy decreases by -150 meV \cite{Spagnoli07c}, so the role of geometrical changes is important. Such a variation (with the reverse sign) has also been produced in other PDA by a uniaxial tension of comparable magnitude, for instance in the PDA pTS \cite{Batchelder78}.
3NPh2 is a typical van der Waals solid, so the volume thermal expansion coefficient should be a little larger than that of 3BCMU which contains hydrogen bonds, but not by much. On the other hand there should be a decrease in repeat unit length comparable to that in poly-4BCMU. The observed exciton energy decreases is -80 meV, barely larger than the expected contribution of $\mid$D$\mid$ implying that the lattice contraction has virtually no effect on the exciton energy. This may possibly be explained by the fact that chains in 3NPh2 are not planar, while blue chains in 3BCMU, 4BCMU or pTS are planar \cite{Enkelmann84, Schott87} and stay so. Upon decreasing T, the chain twist angle may change, and calculations show that the transition energies increase with increasing twist \cite{Filhol09}, so the effects of increased twist and longitudinal contraction may compensate.
\subsection{Luminescence properties}
\begin{figure}
\includegraphics[ width=7cm, height=5cm]{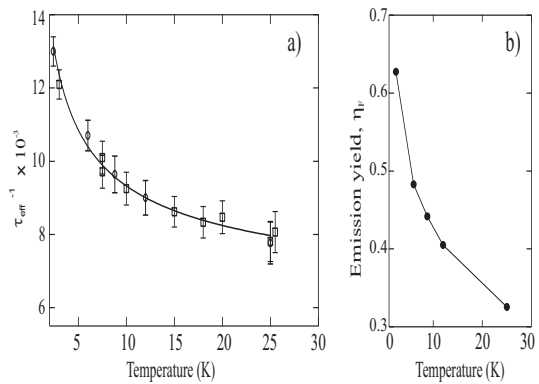} 
\caption{Plot of $\tau_{eff}^{-1}$ versus temperature for the narrow emission (I) at 2.401 eV. Two series of data are plotted (opened circles and squares). The solid line is the result of a fit $\tau_{eff}^{-1}$ = $\tau_{nrad}^{-1}$  +1/b$\sqrt{T}$. b) Calculated quantum yield of emission deduced from the adjustment of $\tau_{rad}$.}
\label{Figure9}
\end{figure}

For two reasons, emissions (I) and (II) have to be associated to different chains population. First the narrow zero-phonon line has its own vibronic progression, characteristic of polydiacetylenes and, second, the temporal study indicates that the radiating states are from different emitters. In the hypothesis of states belonging to the electronic structure of the same chain, one would expect efficient internal conversion from state responsible for emission (I), positioned at higher energy to the state responsible for emission (II), and separated by a few meV. The measured decay time of $\sim$ 82 ps is much too long to account for such process even around 10 K.
As different parts of the crystal are probed, the intensity ratio between peaks (I) and (II) shows only weak variations indicating a relatively homogeneous distribution of each type of chains. It is also remarkable that, if it exists, the Stokes Shift of the different emissions is less than $\sim$ 1 meV.
Due to its perfectly Lorentzian profile, emission (I) was investigated further. Its Lorentzian width, $\Gamma_{I} \sim $1.3 meV at 7 K, places the level of inhomogeneous broadening far below the usual values known in diacetylene mixed crystals. The characteristic time $\hbar / \Gamma_{(I)}$ is two orders of magnitude shorter than the effective lifetime, so it must be associated to a pure dephasing time and leads to T$_{2} \sim$ 0.5 picosecond for this class of chains. Given the model used previously to account for the dephasing time of poly-3BCMU chains (a deformation potential approximation) the coupling to the matrix is slightly larger. 
Given the evolution of the effective lifetime $\tau_{eff}$(T), the radiative time of population (I) was investigated as a function of temperature between 2.4 K and 26 K, following the analysis in \cite{Lecuiller02}; for higher temperatures, the narrow emission cannot be accurately separated from the broader one and the dynamics cannot be distinguished. As seen in figure \ref{Figure9}a, $\tau_{eff}$, clearly increases with T in this temperature range which suggests a similar behaviour for $\tau_{rad}$. Assuming that the non radiative time does not appreciably vary over the studied T range, it is found that $\tau_{rad}$(T) is compatible with a variation proportional to $\sqrt{T}$ as expected for ideal semiconducting wires \cite{Citrin92, Miyauchi09}. We find $\tau_{rad}$ = b$\sqrt{T}$ with b $\sim$ 82$\pm$7 ps·K$^{-1/2}$ and $\tau_{nrad} \sim$ 181$\pm$13 ps by averaging the results of fits from several data series; a typical fit of $\tau_{eff}$ is presented in figure \ref{Figure9}a. The emission quantum yield, $\eta_{F}=\tau_{eff} /\tau_{rad}$, is also deduced. It is a continuously decreasing function of temperature in the considered range. Around 3 K, the analysis provides $\eta_{F} \sim$ 0.6 (figure \ref{Figure9}b) which is, to our knowledge, the highest yield ever reported for a PDA.
The broad emission apparently differs in its behaviour since $\tau_{eff}$ remains a slightly decreasing function of T. In this situation the ratio $\tau_{eff}$/I$_{F}$, I$_{F}$ being the spectrally integrated emission signal (plotted in the inset of figure \ref{Figure6}), was analyzed. The ratio, proportional to $\tau_{rad}$, is still an increasing function of T, however quantitative fit of the data to a square root law, as done for emission (I), does not match, and will require further study.

Red chains of poly-3BCMU and poly-3NPh2 chains of type (I) have thus comparable radiative lifetime at any T \cite{Lecuiller02}. Based on the hypothesis that the calculations of Citrin for a 1D semiconductor are still applicable to the studied quantum wires, $\tau_{rad}$ can be related to the effective mass, m$^{*}_{ex}$, of the exciton \cite{Citrin92}. Assuming equal absorption oscillator strengths for excitons of both chain types, $\tau_{rad}$ is approximately inversely proportional to the product E$_{0}$(m$^{*}_{ex}$)$^{1/2}$. This leads to very similar effective masses (variation of $\sim$10$\%$) in both conjugated structures.

\section{Final remarks}
Including poly-3NPh2, four types of PDA isolated chains having similar degrees of order have now been studied, and their electronic properties characterized: blue poly-4B/3BCMU, red poly-3BCMU, poly-THD and poly-3NPh2 chains. Their zero-phonon transition energy, E$_{0}$, varies at low temperature from 1.8 eV (in poly-4BCMU blue chains) to 2.4 eV (poly-3NPh2), namely an increase of $\sim$30$\%$ whereas the conjugated pattern (bond lengths) is the same. In parallel the chains geometries were determined on the corresponding pure PDA and are seen to continuously vary in the series. Blue chains are found planar \cite{Enkelmann84, Schott87}, a $\theta$=15$^\circ$ twist angle between adjacent monomers was measured in poly-THD \cite{Enkelmann80}, and $\theta$=40$^\circ$ is calculated in the present material \cite{Deschamps}  using a method which also predicts the right geometry for poly-THD \cite{Filhol09}. One should point out that the chain structure is not helical; the angle between the planes containing adjacent monomers takes alternately the +$\theta$ and -$\theta$ values but there is no continuous rotation of the monomers, which would correspond to a constant +$\theta$ angle between them, along the chain axis.

The present work and previous studies clearly demonstrate that E$_{0}$ is strongly affected by the chain conformation via the $\theta$ angle. The impact of $\theta$ variations is not limited to a simple shift of the first optically allowed transition. The strong luminescence of poly-3NPh2 indicates, as in poly-THD \cite{Barisien07}, that the destabilization of the planar structure leads to a reordering of the lowest excited states. For small $\theta$ values 1$^{1}B_{u}$ becomes the lowest singlet excited state, and it is still so for larger $\theta$ values, meaning that, with increasing $\theta$ the dark $A_{g}$ excitonic state shifts towards higher energy faster than $B_{u}$ does. The actual position of the dark $A_{g}$ state in each twisted polymer is unknown and would require specific experiments such as two photon absorption or two photon excited luminescence experiments.
The other feature shared by twisted luminescent PDA wires is the opening of a non radiative channel which becomes active above 50-100 K; it is possible to account for the common luminescence drop by a simple model based on the presence of a quenching energy level slightly above the exciton. If this level is the 2$^{1}A_{g}$ state, a model for 2$^{1}A_{g}$ relaxation bypassing the lower energy 1$^{1}B_{u}$ state is then needed. This might be provided by strong lattice relaxation of the 2$^{1}A_{g}$ state depressing its energy below that of the 1$^{1}B_{u}$ state. Quenching may also be a consequence of singlet fission into two triplet states (see $\S$ IV). Triplet states energies are however not known in those polymers and again, complementary studies and quantum chemical investigations are needed. The available data show similar activation energies for this non radiative channel but the dependence upon $\theta$ is not clear: E$_{act} \sim$ 5 meV in poly-THD, $\sim$ 30-40 meV in poly-3BCMU and $\sim$ 22 meV in poly-3NPh2. The latter result suggests that a finer description of the energetic landscape is necessary. In particular, one may raise the question of the monotonic evolution of the energy difference 1$^{1}B_{u}$ - 2$^{1}A_{g}$ with $\theta$. The effect of a concerted relaxation of other internal coordinates should also be explored.

\section{Conclusion}
Chains resulting from the polymerization of 3NPh2 monomers form highly ordered conjugated systems with a strong luminescence. Two populations of chains coexist in the monomer matrix, one of which at least having emission properties typical of an ideal wire as well as a quantum yield higher than 0.6 at 3K. The chains also have a new geometry far more distorted from the usual planar structure, which adds to the existing set of available structures thus providing an experimental basis for the study of the connection between conformational and electronic properties in model conjugated structures with large electron correlations.
The structure of excitons in perfectly ordered luminescent polydiacetylenic wires can be qualified as robust to variation of the chain conformation: the exciton energy varies and approximately reveals the degree of coupling between the monomers units; however the internal structure and radiative properties are finally not much affected. Comparisons between poly-3NPh2, red poly-3BCMU and poly-THD indicate semi-quantitatively a significant evolution of the different parameters with the angle giving the chain its ``twisted'' conformation.
The outstanding quantum wire properties of isolated red poly-3BCMU chains are shared by other PDA, and we propose that they should be considered as generic of this class of materials. We conjecture that similar properties would be found in other conjugated polymers, if only they could be prepared in the same degree of order.  

\begin{acknowledgements}
Financial support of this work by the Agence Nationale pour la Recherche, France, is gratefully acknowledged (Grant ANR-06-NANO-013). We are indebted to Prof. Daniel Louvard, Head of the Research Department of the Curie Institute (Paris, France), for allowing us to use the IBL 637 for irradiation of the diacetylene compounds and to Mrs Charlotte Bourgeois from INSP for the performed $\gamma$-ray irradiations. We kindly thank Mr. Mathieu Bernard from INSP for technical support in cryogenics.
\end{acknowledgements}

\bibliographystyle{apsrev}
\bibliography{biblio}

\end{document}